\documentclass[final,3p,onecolumn,times]{elsarticle}

\usepackage{amssymb}
\usepackage{amsmath}
\usepackage{epsfig}
\usepackage[active]{srcltx}
\def\slash#1{\settowidth{\bredde}{$#1$}\ifmmode\,\raisebox{.15ex}{/}
\hspace*{-\bredde} #1\else$\,\raisebox{.15ex}{/}\hspace*{-\bredde} #1$\fi}
\def\W #1{\widetilde{#1}}

\def\a{\alpha}
\def\b{\beta}
\newcommand{\bea}{\begin{eqnarray}}
\newcommand{\eea}{\end{eqnarray}}
\newcommand{\bean}{\begin{eqnarray*}}
\newcommand{\eean}{\end{eqnarray*}}
\newcommand{\nn}{\nonumber \\}
\newcommand{\beq}{\begin{equation}}
\newcommand{\eeq}{\end{equation}}
\newcommand{\noi}{\vspace{12pt}\noindent}
\newcommand{\lG}{\raise.3ex\hbox{$\stackrel{\leftarrow}{G}$}}
\newcommand{\lU}{\raise.3ex\hbox{$\stackrel{\leftarrow}{U}$}}
\newcommand{\lP}{\raise.3ex\hbox{$\stackrel{\leftarrow}{{\cal P}}$}}
\newcommand{\leta}{\raise.3ex\hbox{$\stackrel{\leftarrow}{\eta}$}}
\newcommand{\lOmega}{\raise.3ex\hbox{$\stackrel{\leftarrow}{\Omega}$}}
\newcommand{\ldr}{\raise.3ex\hbox{$\stackrel{\leftarrow}{\delta^r}$}}

\def\beqn{\begin{eqnarray}}
\def\eeqn{\end{eqnarray}}

\def\gtwid{\raise.3ex\hbox{$>$\kern-.75em\lower1ex\hbox{$\sim$}}}
\def\ltwid{\raise.3ex\hbox{$<$\kern-.75em\lower1ex\hbox{$\sim$}}}

\journal{}

\begin{document}

\begin{frontmatter}



\title{\huge New Identities among Gauge Theory Amplitudes}


\author{~\\~\\
{  N.E.J. Bjerrum-Bohr$^a$, Poul H. Damgaard$^a$, Bo
Feng$^{b,c}$} and { Thomas S{\o}ndergaard$^a$}\bigskip}

\address{$^a$\small Niels Bohr International Academy and Discovery Center,\\
 \small The Niels Bohr Institute,  \small Blegdamsvej 17, DK-2100
Copenhagen, \small Denmark\bigskip~\\$^b$\small Center of Mathematical Science,\\
\small Zhejiang University, Hangzhou, \small China\\ \bigskip
\small $^c$Kavli Institute for Theoretical Physics China,\\
\small CAS, \small Beijing 100190, China }

\begin{abstract}
Color-ordered amplitudes in gauge theories satisfy non-linear
identities involving amplitude products of different helicity
configurations. We consider the origin of such identities and
connect them to the Kawai-Lewellen-Tye (KLT) relations between
gravity and gauge theory amplitudes. Extensions are made to
one-loop order of the full ${\cal N}=4$ super Yang-Mills
multiplet.
\end{abstract}


\end{frontmatter}


\section*{Introduction}

In a recent paper~\cite{KLTshort} we have proven a series of
non-linear identities among gauge theory amplitudes at tree
level. These identities were discovered accidentally in the
process of giving a field theory proof of the
Kawai-Lewellen-Tye (KLT) relations~\cite{KLT} between gravity
and gauge theory amplitudes. The KLT-relations express
particular combinations of products of two $n$-point
color-ordered gauge theory amplitudes explicitly in terms of an
$n$-point graviton amplitude. It is crucial here that the
helicities in the product match: The graviton helicities $\pm
2$ arise because the corresponding gauge theory helicities are
chosen, correspondingly, as $(\pm 1)\,\times\,(\pm 1)$. There
is thus a direct connection to the factorization properties of
helicities of massless external states. What happens if
helicities do not match? In ref.~\cite{KLTshort} we proved for
all $n$ that if the helicity of just one leg is flipped, the
particular combinations of products that enter the right hand
side of the KLT-relations {\em vanish}. Viewed from the gauge
theory side this gives a series of quite unusual identities
among color-ordered amplitudes.

\noi One of the surprises in amplitude calculations in recent
years was the discovery by Bern, Carrasco and Johansson (BCJ)
of a new set of linear relations among color-ordered gauge
theory amplitudes~\cite{BCJ} (see also the extension to scalar
and fermionic matter in~\cite{Sondergaard}). These identities
were shown to follow also from the field theory limit of string
theory~\cite{BDV,Stieberger} and in ref.~\cite{BDV} it was
proven that this implies that the minimal basis of color-order
gauge theory amplitudes is of size $(n-3)!$, rather than the
$(n-2)!$ that would be inferred from just photon decoupling and
Kleis-Kuijff relations~\cite{KK,Lance}. Very recently, a purely
field theoretic proof of all these identities has been
presented~\cite{Feng}, see also~\cite{Jia}. An alternative
understanding of the BCJ-relations has also been achieved
through a study of the generalized Jacobi-like identities they
imply for residues of poles~\cite{many}. Squaring-relations of
these pole structures yield alternative forms of KLT-relations.
As shown recently by Tye and Zhang in the second paper of
ref.~\cite{many}, this actually follows from the field theory
limit of the heterotic string. Such squaring relations even
seem to hold beyond tree level~\cite{Bern:2010}.

\noi While the new identities we present in this paper do not
have the power of BCJ-relations, and in particular by necessity
cannot reduce the $(n-3)!$ growth of the amplitude basis, the
identities themselves are so striking and unusual that they
deserve further study. As we will show below, some of the
identities carry over to at least one-loop level. All of these
identities are on the gauge theory side alone, with no relation
to gravity. Nevertheless, an understanding of these identities
from the point of view of string theory could be most
interesting.

\section*{New form of KLT and gauge theory identities}

We will here briefly review the results of ref.~\cite{KLTshort}
and phrase them in a new form which is more practical. Let
$A_n(1,2,\ldots,n)$ and $\widetilde{A}_n(1,2,\ldots,n)$ denote
$n$-point color-ordered gauge theory amplitudes of fixed
helicity. Furthermore, let
\begin{equation}
{\cal S}[i_1,\ldots,i_k|j_1,\ldots,j_k]_{p_1} \equiv
\prod_{t=1}^k \big(s_{i_t 1}+\sum_{q>t}^k \theta(i_t,i_q)
s_{i_t i_q}\big) \label{Sdef}\,,
\end{equation}
where $s_{12\ldots i} \equiv (p_1 + \ldots + p_{i})^2$
and $\theta(i_a,i_b)$ is zero if $i_a$ sequentially comes
before $i_b$ in $\{j_1,\ldots,j_k\}$. Otherwise it is unity.
The function ${\cal S}$ has the following symmetry ${\cal
S}[i_1,\ldots,i_k|j_1,\ldots,j_k]  = {\cal
S}[j_k,\ldots,j_1|i_k,\ldots,i_1]\, .\label{Ssym}$
We also introduce a dual $\widetilde{\mathcal{S}}$ defined by
\begin{align}
\widetilde{\mathcal{S}}[i_1,\ldots,i_{k}|j_1,\ldots,j_{k}]_{p_n} \equiv
\prod_{t=1}^{k} \big(s_{j_t n}+\sum_{q<t} \theta(j_q,j_t)
s_{j_t j_q}\big) \label{dualSdef}\,,
\end{align}
where again $\theta(j_a,j_b)$ is zero if $j_a$ sequentially comes
before $j_b$ in $\{i_1,\ldots,i_{k}\}$. Otherwise it is unity.

\noi A main result of~\cite{KLTshort} was the proof of the
following equation (we will throughout use the shorthand
notation: $\gamma_{2,n-1}$ for the ordering of legs
$2,3,\ldots,n-1$ in the amplitude ${\widetilde{A}_n}$ and
likewise $\beta_{2,n-1}$ for the order in $A_n$).
\begin{align} X_n^{(n_+,n_-)} = \displaystyle
\sum_{\gamma,\beta}{\widetilde{A}_n(n,\gamma_{2,n-1},1) {\cal S}[
\gamma_{2,n-1}|\beta_{2,n-1}]_{p_1} A_n(1,\beta_{2,n-1},n)\over
s_{12\ldots (n-1)}}\label{vanish1}\,,
\end{align}
where $n_+$ ($n_-$) denotes the number of positive (negative)
helicity legs in $A_n$ which is changed to negative (positive)
helicity legs in $\widetilde{A}_n$ and we sum over all
permutations of legs in sets $\gamma$ and $\beta$. When no
helicities are changed we obtain the gravity amplitude, {\it
i.e.} $X_n^{(0,0)}=(-1)^n M_n(1,2,\ldots,n)$ with
$M_n(1,2,\ldots,n)$ denoting the $n$-point gravity amplitude.
However, just by flipping one helicity on one of the two gluon
amplitudes we have $X_n^{(1,0)} =X_n^{(0,1)} = 0$. These are
quite surprising identities for gluon amplitudes of definite
helicities.

\noi On-shell, the identities implied by (\ref{vanish1}) have
both vanishing numerator and denominator, and are therefore
ill-defined. However, as explained in ref.~\cite{KLTshort}, one
should understand the expression in terms of a regularization
that takes the $n$'th leg off-shell. In detail, this can be
achieved, for instance, by shifting momenta as follows: $p_1
\to p_1 - x q\,,$ and $p_n \to p_n + x q\,.$
Here $x$ is an
arbitrary parameter and $q^2 = 0 = p_1\cdot q$ but $p_n\cdot q
\neq 0$. This clearly preserves overall energy-momentum
conservation, keeps the external leg 1 on-shell, but makes
$p_n^2 = s_{12\ldots (n-\!1)} \neq 0$. The expression
(\ref{vanish1}) is then well-defined, and one obtains the
correct result as the limit of $x\rightarrow 0$. How to
systematically take this limit was explained in~\cite{KLTshort}
(see also \cite{BDFS}) .

\noi Although eq.~\eqref{vanish1} has the advantage of being
manifestly symmetric in $(n-2)!$ legs, it is for practical
purposes more convenient to use an equivalent form~\cite{BDFS}
which has fewer terms and does not require regularization. Such
a general form is given by
\begin{align}
X_n^{(n_+,n_-)}  = & -\sum_{\sigma\in S_{n-3}}\sum_{\W\a\in
S_{j-2}}\sum_{\b\in S_{n-1-j}}
  \W A(\W\a(\sigma_{2,j-1}),1,n-1,\W\b(\sigma_{j,n-2}),n)  {\cal S}[
\W\a(\sigma_{2,j-1})|\sigma_{2,j-1}]_{p_1}\nn
&  \hspace{4cm}\times {\cal \W S}[\sigma_{j,n-2}
|\W \b(\sigma_{j,n-2})]_{p_{n-1}}A(1,\sigma_{2,j-1},\sigma_{j,n-2}, n-1,n)\label{KLT-new-J}\,,
\end{align}

\noi
The particular KLT-expression that was conjectured in ref.~\cite{Bern:1998} is
equivalent to (\ref{KLT-new-J}) in the special case of $j=[n/2-1]$.
However, eq.~\eqref{KLT-new-J} is more general and valid
for any $j$. Particularly interesting forms arise when we
take either
the left or the right $j$-set empty. Then we get
two highly symmetric relations:
\begin{equation}
X_n^{(n_+,n_-)}  =  -\sum_{\gamma,\beta\in S_{n-3}} \W
A(n-1,n,\gamma_{2,n-2},1) {\cal
S}[\gamma_{2,n-2}|\beta_{2,n-2}]_{p_1} A(1,\beta_{2,n-2}, n-1,n)
\,,\label{KLT-Pure-S}
\end{equation}
and
\begin{equation}
X_n^{(n_+,n_-)}  =  -\sum_{\gamma,\beta\in S_{n-3}}
A(1,\beta_{2,n-2}, n-1,n) {\cal \W S}[
\beta_{2,n-2}|\gamma_{2,n-2}]_{p_{n-1}} \W A(1,n-1,\gamma_{2,n-2},n)
\,.\label{KLT-Dual-S}
\end{equation}
It is interesting to observe how eq.~\eqref{KLT-Pure-S}
resembles the numerator of eq.~\eqref{vanish1}. The difference
lies only in the number of legs being permuted. Note that both
of the two gauge amplitudes are expanded in a minimal basis of
$(n-3)!$ amplitudes. However, the basis of amplitudes for $A$
is not the same as that of $\widetilde{A}$. This is a very
simple form of the KLT-relations and the new gauge theory
identities.

\section*{Flipping several helicities}

As noted, one can generate new
identities among gauge theory amplitudes by simply
flipping the helicity of
one leg in the $n$-point KLT relation. In this section
we consider the cases with more than one
flipped helicity. We begin by writing out some explicit
examples in the form of eq.~\eqref{KLT-Pure-S}.

\noi
In the 4-point case $\mathcal{S}[2|2] = s_{12}$ so that
\begin{align}
X_4^{(n_+,n_-)} = -s_{12}A_4(1,2,3,4)\widetilde{A}_4(3,4,2,1).
\end{align}
which give trivial zeros when $n_+\neq n_-$. For example, with
$(n_+,n_-)=(0,1)$ we get
\begin{align}
0 = -s_{12}A_4(1^-,2^-,3^+,4^+)\widetilde{A}_4(3^+,4^+,2^+,1^-).
\end{align}
which expresses nothing but the standard MHV helicity selection rule.

\noi
In the 5-point case we have
\begin{align}
X_5^{(n_+,n_-)} ={}&
-s_{12}A_5(1,2,3,4,5)\big[ s_{13}\widetilde{A}_5(4,5,2,3,1)
+ (s_{13}+s_{23})\widetilde{A}_5(4,5,3,2,1)\big] \nonumber \\
& - s_{13}A_5(1,3,2,4,5)\big[s_{12}\widetilde{A}_5(4,5,3,2,1) + (s_{12}
+s_{23})\widetilde{A}_5(4,5,2,3,1)\big] .
\end{align}
Explicit calculations show that we get zeros in cases like
(\textit{e.g.} with $(n_+,n_-)=(1,0)$)
\begin{align}
0 ={}& s_{12}A_5(1^-,2^-,3^+,4^+,5^+)
\big[ s_{13}\widetilde{A}_5(4^+,5^+,2^-,3^-,1^-)
+ (s_{13}+s_{23})\widetilde{A}_5(4^+,5^+,3^-,2^-,1^-)\big] \nonumber \\
& + s_{13}A_5(1^-,3^+,2^-,4^+,5^+)
\big[ s_{12}\widetilde{A}_5(4^+,5^+,3^-,2^-,1^-)
+ (s_{12}+s_{23})\widetilde{A}_5(4^+,5^+,2^-,3^-,1^-)\big] .
\end{align}
In contrast to the 4-point case, this is already a new non-trivial
identity. We also get zero when we do helicity flips in the category
of $(n_+,n_-)=(2,1)$, $(n_+,n_-)=(2,0)$, or $(n_+,n_-)=(3,2)$.

\noi
Finally, we give the explicit expression for the 6-point case,
\begin{align}
X_6^{(n_+,n_-)} ={}& -s_{12}s_{13}A_6(1,2,3,4,5,6)
\big[ s_{14}\widetilde{A}_6(5,6,2,3,4,1)
+ (s_{14}+s_{34})\widetilde{A}_6(5,6,2,4,3,1) \nonumber \\
&\hspace{6.5cm}+(s_{14}+s_{34}+s_{24})
\widetilde{A}_6(5,6,4,2,3,1)\big] \nonumber \\
& -s_{12}(s_{13}+s_{23})A_6(1,2,3,4,5,6)
\big[ s_{14}\widetilde{A}_6(5,6,3,2,4,1)
+ (s_{14}+s_{24})\widetilde{A}_6(5,6,3,4,2,1) \nonumber \\
&\hspace{8cm} + (s_{14}+s_{24}+s_{34})
\widetilde{A}_6(5,6,4,3,2,1) \big] \nonumber \\
& + \mathcal{P}(2,3,4), \label{6ptex}
\end{align}
where we similarly get vanishing relations
in all non-trivial cases where $n_+\neq n_-$. From these simple examples
we seem to extract the following general rule:
\begin{align}
n_+ \neq n_- \quad \Longrightarrow \quad  X_n^{(n_+,n_-)}=0.\label{rule}
\end{align}
This is what we will show below.

\section*{Proof of new gauge theory identities with several
flipped helicities}

To simplify the proof of (\ref{rule}) we use the form of
eq.~\eqref{vanish1}. The reason is that eq.~\eqref{vanish1} is
manifestly symmetric in $(n-2)!$ legs. Choosing the two
remaining legs as 1 and $n$, this is ideally suited for a proof
based on BCFW-recursion~\cite{BCFW}.

\noi We do the proof by induction. We thus assume that we have
verified the rule (\ref{rule}) for the identities up to $n-1$
points and now we want to show that this implies the rule at
$n$ points. We thus look at the $n$-point identity and imagine
having changed $n_+$ of the positive helicity legs and $n_-
\neq n_+$ of the negative helicity legs in $\widetilde{A}_n$
(compared to $A_n$). Doing a BCFW-shift in the legs $1$ and
$n$, we can then consider following contour integral
(${C_\infty=0}$),
\begin{align} 0 =
\oint \frac{dz}{z}X_n^{(n_+,n_-)}(z) = X_n^{(n_+,n_-)}(0) +
(\mathrm{residues\:\:for}\:\:z\neq 0)\,.
\end{align}
We treat separately the following two classes of contributions
(for each residue):\medskip
\begin{itemize}
\item[(A)] We have a pole appearing in only one of the
    amplitudes $\widetilde{A}_n$ or $A_n$.
\item[(B)] We have a pole that is present in both
    amplitudes $\widetilde{A}_n$ and $A_n$.
\end{itemize}\medskip
Starting with case (A), we first note that one needs to consider only
$\widetilde{A}_n$ having the pole. (If the pole
instead sits in $A_n$ there is a similar argument, by symmetry).

\noi
Looking back at eq.~(\ref{vanish1}), we can compute the residue
of the pole $s_{\widehat{1}2..k}$ as $-\lim_{z\rightarrow
z_{12..k}} \big[ s_{\widehat{1}2..k}(z) X_n^{(n_+,n_-)}(z)
\big]/s_{12..k}$, where $z_{12..k}$ is the $z$-value that makes
$s_{\widehat{1}2..k}$ go on-shell. From this we get
\begin{align}\label{partA}
&\frac{(-1)}{s_{\widehat{1}2..n-1}} \sum_{\gamma,\sigma,\beta}
\frac{\sum_h\widetilde{A}_{n-k+1}(\widehat{n},\gamma,-\widehat{P}^h)
\widetilde{A}_{k+1}(\widehat{P}^{-h},\sigma,\widehat{1})}{s_{12..k}}
\times \mathcal{S}[\gamma\sigma|\beta_{2,n-1}]_{p_1}
A_n(\widehat{1},\beta_{2,n-1},\widehat{n})\,,
\end{align}
where we can rewrite $\mathcal{S}[\gamma\sigma|\beta_{2,n-1}]_{p_1}$ as
$\mathcal{S}[\sigma|\rho_{2,k}]_{p_1}\! \times\! \text{(something
independent of } \sigma)$. Here $\rho_{2,k}$ stands for the
relative ordering of legs $2,3,\ldots,k$ in $\beta$. We thus
conclude that eq.~\eqref{partA} is identically zero
(at $z=z_{12..k}$) since
\begin{align}
\sum_{\sigma}\widetilde{A}_{k+1}(\widehat{P}^{-h},\sigma,\widehat{1})
\mathcal{S}[\sigma|\rho_{2,k}]_{p_1} = 0\,,
\end{align}
independently of the helicity configuration in
$\widetilde{A}_{k+1}(\widehat{P}^{-h},\sigma,\widehat{1})$. This
holds also with the regularization discussed above. All terms coming
from the class (A) above will therefore not contribute to any
residues.

\noi
For the class (B), consider again eq.~(\ref{vanish1}) and
the $s_{12..k}$ pole contribution which is now present in both
$\widetilde{A}$ and $A$. The residue takes the form
\begin{align}
\frac{(-1)}{s_{\widehat{1}2...(n-1)}}
\sum_{\gamma,\beta,\sigma,\alpha} \left[ \frac{\sum_h
\widetilde{A}(\widehat{n},\gamma,\widehat{P}^{-h})
\widetilde{A}(-\widehat{P}^{h},\sigma,\widehat{1})}{s_{12..k}}
\right]
 \mathcal{S}[\gamma\sigma|\alpha\beta]_{p_1}
\left[ \frac{\sum_h A(\widehat{1},\alpha,-\widehat{P}^h)
A(\widehat{P}^{-h},\beta,\widehat{n})}
{s_{\widehat{1}2..k}} \right]\,, \label{bothpoles}
\end{align}
where the sets $\beta \equiv \beta_{k+1,n-1}$ and $\alpha
\equiv \alpha_{2,k}$ and the subscript indices on $A$ and
$\widetilde{A}$ have been suppressed for clarity. In (\ref{bothpoles})
above one of the shifted $s_{\widehat{1}2..k}$ poles has been
substituted by an unshifted pole $s_{12..k}$ from calculating
the single-pole residues. Noting that $s_{\widehat{1}2..n-1} =
s_{\widehat{P}k+1..n-1}$, and using
$\mathcal{S}[\gamma\sigma|\alpha\beta]_{p_1}
=\mathcal{S}[\sigma|\alpha]_{p_1} \times
\mathcal{S}[\gamma|\beta]_{\widehat{P}}$, we can rewrite the
above expression
\begin{align}
&\frac{1}{s_{12..k}} \sum_h\left[ \Bigg(\sum_{\sigma,\alpha}
\frac{\widetilde{A}(-\widehat{P}^{h},\sigma,\widehat{1})\mathcal{S}
[\sigma|\alpha]_{p_1}A(\widehat{1},\alpha\,,-\widehat{P}^
{h\ })}
{s_{\widehat{1}2..k}} \Bigg)
\Bigg(\sum_{\gamma,\beta}
\frac{\widetilde{A}(\widehat{n},\gamma,\widehat{P}^{-h})
\mathcal{S}[\gamma|\beta]_{\widehat{P}}
A(\widehat{P}^{-h},\beta,\widehat{n})}{s_{\widehat{P}k+1..(n-1)}}
\Bigg) \right] \label{partB1} \\
&\hspace{-0.37cm}+\frac{1}{s_{12..k}} \sum_h\left[ \Bigg(\sum_{\sigma,\alpha}
\frac{\widetilde{A}(-\widehat{P}^{h},\sigma,\widehat{1})\mathcal{S}
[\sigma|\alpha]_{p_1}A(\widehat{1},\alpha,-\widehat{P}^{-h})}
{s_{\widehat{1}2..k}} \Bigg)
\Bigg(\sum_{\gamma,\beta}
\frac{\widetilde{A}(\widehat{n},\gamma,\widehat{P}^{-h})
\mathcal{S}[\gamma|\beta]_{\widehat{P}}
A(\widehat{P}^{h\  },\, \beta,\widehat{n})}{s_{\widehat{P}k+1..(n-1)}}
\Bigg) \right]. \nonumber
\end{align}
We now want to argue that line one and two of~\eqref{partB1}
will always be zero separately. We start by giving the
arguments mostly in words, and afterward give another, more
mathematical, proof. In what follows when we talk about the set
$\alpha$ we will actually mean the above $\alpha$ set plus leg
1, and when talking about the $\beta$ set we mean the above
$\beta$ set plus leg $n$. There are several different cases.

\begin{enumerate}
\item{All flipped legs lie within the $\alpha$ set
    in~\eqref{partB1}. In this case the first term in line
    one is zero by the induction principle, and (at least)
    the second term in line two is zero (since it will be a
    single-leg flip relation).\smallskip}
\item{ There are flipped legs in both the $\alpha$ and
    $\beta$ set in~\eqref{partB1}, but the legs in the
    $\alpha$ set still satisfy the rule. In this case the
    first term in line one is again zero because of the
    induction assumption. In line two we can have two kind
    of situations:\smallskip} \subitem{a) The difference in
the number of negative and positive helicity-flipped legs
in the $\alpha$ set is greater than 1, so the addition of
one flipped leg (\textit{i.e.} $-\widehat{P}$) does not
ruin the zero.\smallskip} \subitem{b) The difference in the
number of negative and positive helicity-flipped legs in
the $\alpha$ set is equal to 1. Here the first term of line
two will only be zero for one of the $h=\pm$ values. But
since the $\widehat{P}$ leg in the amplitude containing the
$\beta$ set always has the opposite helicity of the
$-\widehat{P}$ leg in the amplitude containing the $\alpha$
set, the relation with the $\beta$ set must be zero in the
case where the relation with the $\alpha$ set is not
necessarily zero. If this was not so, we would have chosen
$n_+=n_-$ which is against our starting
assumption.\smallskip}
\item{There are flipped legs in both the $\alpha$ and
    $\beta$ sets in~\eqref{partB1}, but there is an equal
    number of negative and positive helicity-flipped legs
    in the $\alpha$ set. However, since $n_+\neq n_-$ the
    remaining flipped legs in the $\beta$ set must now
    satisfy our rule, and therefore make the second term of
    line one vanish. Since the $\alpha$ set contains an
    equal number of flipped legs, adding one flipped leg,
    either of positive or negative helicity, will give us
    an unequal number, and the first term of line two
    therefore vanishes.}
\item{All flipped legs lie in the $\beta$ set
    in~\eqref{partB1}. These contributions are zero by
    arguments similar to those in 1).}
\end{enumerate}

The above result can also be stated in a more mathematical
language. In eq.~\eqref{partB1}, we see that the counting of
$n_+, n_-$ in line one schematically can be written as
\bea (n_+, n_-)\to (n_+^{\a},n_-^\a)\times
(n_+^\b,n_-^\b)\,,\eea
while line two  can be written as
\bea (n_+, n_-)\to (n_+^{\a}+1,n_-^\a)\times (n_+^\b,n_-^\b+1)+
(n_+^{\a},n_-^\a+1)\times (n_+^\b+1,n_-^\b)\,.\eea
For line one  to be nonzero, we need to have
 \bea n_+^{\a}=n_-^\a, \qquad\mathrm{and} \qquad n_+^\b=n_-^\b, \qquad \Longrightarrow \qquad n_+=n_-\,,\eea
which is not true. For line two to be nonzero, we need to have
\bea n_+^{\a}+1=n_-^\a,\qquad\mathrm{and} \qquad
n_+^\b=n_-^\b+1, \qquad \Longrightarrow \qquad n_+=n_-\,,\eea
or
\bea n_+^{\a}=n_-^\a+1,\qquad\mathrm{and} \qquad
n_+^\b+1=n_-^\b, \qquad \Longrightarrow \qquad n_+=n_-\,,\eea
which is again not true from our start assumption.

This concludes the induction proof of (\ref{rule}).

\section*{Identities at one-loop level}

So far our discussion has been restricted to tree level. Identities
among tree level amplitudes can clearly have consequences for loop
amplitudes through generalized unitarity~\cite{generalizeduni}, and
in particular quadruple-cut constructions~\cite{qcuts}. Indeed, we
find a series of new identities also at one-loop level of gauge
theories. To illustrate this, we first remind the reader of some of
the one-loop results derived in ref.~\cite{BBSwansea}. In that paper
expressions for ${\cal N}=8$ supergravity box-coefficients were
linked directly to products of box coefficients of ${\cal N}=4$
super Yang-Mills. Box coefficients are multiplying a class of
integral functions known as scalar box integrals. Such integrals
have four propagator lines which are integrated over. As was shown
in ref.~\cite{BBSwansea}, one can derive the following set of
relations valid at six-point scattering
\begin{equation}
\begin{split}
 c_{N=8}^{(ab)c(de)f}
&= 0\,, \\
 c_{N=8}^{(ab)(cd)ef}  &=
 2 s_{ab}s_{cd} \times \biggl( \sum
 c_{i}^{(ab)(cd)ef}
\times
 c_{i}^{(ba)(dc)ef} \biggr)\,,
\cr
 c_{N=8}^{(abc)def} & =\null  2 s_{ab}s_{c\ell_c} \sum
\biggl( c_{i}^{(abc)def}
 c_{i}^{(bac)def}
+ c_{i}^{(abc)def} c_{i}^{(bca)def}
+ c_{i}^{(abc)def}  c_{i}^{(cba)def}
\biggr)
\cr
&
\; \; + 2 s_{ac}s_{b\ell_c} \sum
\biggl(  c_{i}^{(acb)def}  c_{i}^{(cab)def}
+ c_{i}^{(acb)def} c_{i}^{(cba)def}
+ c_{i}^{(acb)def} c_{i}^{(bca)def}
\biggr)\,,
\end{split}
\end{equation}
for all choices of helicities. We refer to
ref.~\cite{BBSwansea} for the precise definition of the box
coefficients $c_{N=8}$ and $c_i$ that enter the above
expression.

\noi Using our rule for vanishing identities between gauge
theory amplitudes this leads to results such as (the
superscript $\pm$ on leg $c$ denotes that helicity on that leg
has been chosen oppositely)
\begin{equation}\begin{split}
0 &=  2
s_{ab}s_{c\ell_c} \sum \biggl(
c_{i}^{(abc^+)def}  c_{i}^{(bac^-)def} +
c_{i}^{(abc^+)def} c_{i}^{(bc^-a)def} + c_{i}^{(abc^+)def}
 c_{i}^{(c^-ba)def} \biggr)
\\& + 2 s_{ac}s_{b\ell_c} \sum
\biggl(  c_{i}^{(ac^+b)def}  c_{i}^{(c^-ab)def}
+ c_{i}^{(ac^+b)def} c_{i}^{(c^-ba)def}
+ c_{i}^{(ac^+b)def} c_{i}^{(bc^-a)def}
\biggr)\,.
\end{split}\end{equation}
{}From this one can directly link coefficients of different
boxes with opposite helicity configurations. This provides a
link at one-loop level for our gauge theory identities. One
readily checks that they indeed are satisfied. We could of
course consider flipping more legs and also higher $n$-point
functions in a similar fashion. This as well as multi-loop
considerations are beyond the scope of this paper and we leave
it for future work.

\section*{On the origin of the identities}

As we have seen above, when $n_+=n_-=0$ our formulae simply
express the standard KLT-relations, \textit{i.e.} $X_n^{(0,0)}$
give the gravity tree-level amplitude, but when $n_+\neq n_-$
we get $X_n^{(n_+,n_-)}=0$, the new gauge theory identities. It
is interesting to consider what happens if $n_+=n_-\neq 0$. In
that case we do not get identities among gauge theory
amplitudes. Since our focus in this paper is just on these
gauge theory identities, our discussion around this case will
be brief.

\noi The origin of the non-vanishing of $X_n^{(n_+,n_-)}$ in
the cases where $n_+=n_-\neq 0$ can be illustrated by a
rewriting of the 4-point case with $n_+=n_-=1$. We have,
\begin{align}
-s_{12}A_4(1^-,2^-,3^+,4^+)\widetilde{A}_4(3^-,4^+,2^+,1^-) =
 -s_{12}A_4(1^-,2_s,3_s,4^+)\widetilde{A}_4(3_s,4^+,2_s,1^-),
\end{align}
where the $s$ subscript denotes a scalar particle, and where we
have used supersymmetric Ward identities to write the purely
gluonic amplitudes in terms of gluon/scalar amplitudes. We see
that the $n_+=n_-=1$ case is nothing but a KLT-relation
involving scalars, and therefore obviously should not vanish.
Doing the same in the case of $n_+=n_-=2$ we obtain 4-point
scalar amplitudes. One could off course also think of more
exotic relations involving for example flipped fermion spins
(see e.g. refs.~\cite{Bern:1999bx} for different types of mixed
matter relations for which spin flips could be considered).
In~\cite{Bianchi:2008pu} different relations are considered
between operators of ${\cal N} = 8$ and ${\cal N} =4$ it would
also be interesting to think of flips of spins in such a
context.

\noi
Because of the simple form of supersymmetric Ward identities for
MHV amplitudes,
the above rewriting can easily be extended to higher-point relations.
It is straightforward
to check that it holds in the 5-point case as well.
It thus becomes clear that the new gauge theory identities arise when there
are no matching amplitudes in the (super)gravity sector: These are the
``forbidden'' combinations that provide constraints on the gauge theory
sectors alone. An understanding of this helicity selection rule
at the string theory level could be most interesting.

\section*{Conclusions}

In this paper we have discussed some rather intriguing
identities among Yang-Mills amplitudes. They were found using
inspiration from the field theory limit of the well-known
KLT-relations after flipping one or more helicities between the
two gauge theory amplitudes on the right hand side. We have
found it useful to present our results using a generalized and
more symmetric version of the KLT-relations which we have
uncovered in the process of this investigation. Although we
have only considered gluon scattering amplitudes in this paper,
extensions certainly exist for amplitudes involving external
fermions and scalars. We have also considered the impact of
these identities on loop amplitudes. Relations between box
coefficients of one-loop amplitudes in ${\cal N}=4$ super
Yang-Mills theory follow in this way. Further progress in this
direction is likely and it would clearly be interesting to
investigate more identities at one and possibly multi-loop
level.  It would also be an interesting task to understand
better the origin of these identities. It seems quite likely
that string theory also here holds the clue. This seems a
promising avenue for future studies.

\subsection*{Acknowledgements}
(BF) would like to acknowledge funding from Qiu-Shi, the
Fundamental Research Funds for the Central Universities with
contract number 2009QNA3015, as well as Chinese NSF funding
under contract No.10875104. (BF) would also like to thank the
NBIA, where a major part of this work was done, for
hospitality.

\end{document}